\documentclass[aps, floatfix, prl, longbibliography,amsmath,amssymb,twocolumn]{revtex4-1}
\usepackage{graphicx,xcolor,dcolumn,xcolor}
\usepackage{bm}
\newcommand{\bra}[1]{\langle #1\rangle}
\def \V {\mathcal{V}}

\def \uu {\bm{u}}
\def \Rey  {\mbox{Re}}
\def \Ca  {\mbox{Ca}}
\def \gdot {\dot{\gamma}}
\def \muf {\mu_{\rm f}}
\def \sigmaf {\sigma_{\rm f}}
\def \sigmas {\sigma_{\rm s}}
\def \rhof {\rho_{\rm f}}
\def \Tay {\mathcal{T}}
\def \mus {\mu_{\rm s}}
\def \phim {\Phi_{\rm m}}
\def \phie {\Phi_{\rm e}}

\newcommand{\Eq}[1]{Eq.~(\ref{#1})}
\newcommand{\Fig}[1]{Fig.~(\ref{#1})}
\newcommand{\subfig}[2]{Fig.~(\ref{#1}#2)}
\newcommand{\etal}{et al.\ }
\newcommand{\ie}{i.e.,\ }

\definecolor{cinnamon}{rgb}{0.82, 0.41, 0.12}

\newcommand{\beginsupplement}{%
        \setcounter{table}{0}
        \renewcommand{\thetable}{S\arabic{table}}%
        \setcounter{figure}{0}
        \renewcommand{\thefigure}{S\arabic{figure}}%
     }
     
\begin{document}
\title{Rheology of suspensions of viscoelastic spheres:\\deformability as an effective volume fraction}
\author{Marco E. Rosti$^1$}
\author{Luca Brandt$^1$}
\author{Dhrubaditya Mitra$^2$}
\affiliation{$^1$ Linn\'{e} Flow Centre and SeRC (Swedish e-Science Research Centre), \\KTH Mechanics, SE 100 44 Stockholm, Sweden\\$^2$ NORDITA, Royal Institute of Technology and Stockholm University,\\Roslagstullsbacken 23, SE 106 91 Stockholm, Sweden}

\date{\today}

\begin{abstract}
We study suspensions of {\it deformable} (viscoelastic) spheres in a Newtonian solvent in plane Couette geometry, by means of direct numerical simulations. We find that in the limit of vanishing inertia the  effective viscosity~$\mu$ of the suspension increases as the volume-fraction occupied by the spheres $\Phi$ increases and decreases as the elastic modulus of the spheres $G$ decreases; the function $\mu(\Phi,G)$ collapses to an universal function, $\mu(\phie)$, with a reduced effective volume fraction $\phie(\Phi,G)$. Remarkably, the function $\mu(\phie)$ is the well-known Eilers fit that describes the rheology of suspension of {\it rigid} spheres at all $\Phi$. Our results suggest new ways to interpret macro-rheology of blood. 
\end{abstract}

\maketitle
Most of the fluids we encounter in our everyday life -- from the mud we wade through to the blood that flows through our veins -- are complex fluids. One of the most useful ways to understand the rheology of complex fluids is to model them as suspensions of objects in a Newtonian solvent with dynamic viscosity $\muf$ and density $\rhof$~\cite{stickel_powell_2005a, mewis_wagner_2012a}. The rheology of suspensions can be quite complex, as it depends on the shear-rate $\gdot$, the volume-fraction $\Phi$ occupied by the suspended objects, the properties of the suspended objects themselves (some examples are rigid spheres, bubbles, a different fluid enclosed in a membrane), and their poly-dispersity. In the simplest case of rigid spheres in the limit of small $\Phi$, and  vanishing inertia (small $\gdot$), also ignoring thermal fluctuations~(infinite Peclet number), the fractional increase in the effective viscosity of the suspension is given by~\cite[see, e.g.,][section 4.11]{batchelor_2000a}
\begin{equation}
\frac{\mu}{\muf} = 1 + \frac{5}{2}\Phi + \mathcal{O}(\Phi^2) \/,  
\label{eq:mu}
\end{equation}
At present there is no theory that allows us to calculate $\mu$ for any given $\Phi$ and $\gdot$.  Different empirical formulas provide a good description to the existing experimental and numerical results~\cite{ferrini_ercolani_de-cindio_nicodemo_nicolais_ranaudo_1979a, zarraga_hill_leighton-jr_2000a, singh_nott_2003a, kulkarni_morris_2008a}. Among those, we consider here the Eilers formula~\cite{stickel_powell_2005a, mewis_wagner_2012a},
\begin{equation}
\frac{\mu}{\muf} = \left[ 1 + B \frac{\Phi}{1-\Phi/\phim}\right]^2 \/,
\label{eq:Eiler}
\end{equation}
which fits well the experimental and numerical data~\cite{zarraga_hill_leighton-jr_2000a,singh_nott_2003a} for both low and high values of $\Phi$, up to about $0.6$. In the expression above, $\phim$ is the geometrical maximum packing fraction, and $B$ is a constant, and the  best fit to the data yields $\phim=0.58 - 0.63$ and $B=1.25 - 1.7$. If the radius $R$ of the spheres and the shear-rate are large enough, the particle Reynolds number, defined as $\Rey \equiv (\rhof R^2\gdot)/\muf$, is greater than unity, inertial effects are non negligible and the viscosity $\mu=\mu(\Phi,\Rey)$. Remarkably, direct numerical simulations (DNS) in Ref.~\cite{picano_breugem_mitra_brandt_2013a} demonstrated that the Eilers fit is a good approximation even for inertial suspensions if $\Phi$ in \Eq{eq:Eiler} is replaced by an {\it increased effective volume fraction}, $\phie(\Phi,\Rey)$. Due to the increase of the effective volume fraction with the applied shear, the suspension viscosity increases, a phenomenon called  inertial shear-thickening. 

In this letter we add a different complexity to this problem, one that  is particularly important to understand rheology of biological flows; while keeping small $\Rey$, we allow the suspended particles to be {\it deformable}. In particular, we model the spheres as viscoelastic material with an elastic shear--modulus $G$ and viscosity $\mus$. Thereby we introduce two new dimensionless parameters: the Capillary number $\Ca\equiv \muf\gdot/G$ and the viscosity ratio $K\equiv \mus/\muf$.  This problem has a long history starting with the work by Taylor~\cite{taylor_1932a} who assumed small deformation ($\Ca \to 0$) and showed that for small $\Phi$ the coefficient of the linear term on the right-hand-side of \Eq{eq:mu} is $(5K+2)/(2K+2)$. Later analytical calculations~\cite{cox_1969a, frankel_acrivos_1970a, choi_schowalter_1975a, pal_2003a, gao_hu_castaneda_2012a} attempted to extend the result of Taylor to higher order in $\Phi$ and $\Ca$ using perturbative expansions. Recently, numerical simulations~\cite{matsunaga_imai_yamaguchi_ishikawa_2016a} have been used to estimate the deformation and suspension viscosity for elastic capsules.

\begin{figure}[]
  \centering
  \includegraphics[width=0.45\textwidth]{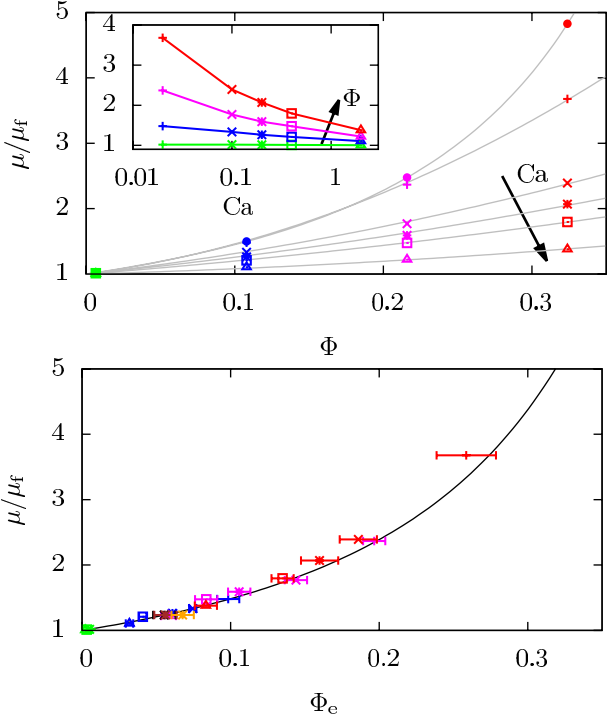}
  \caption{ (top) The fractional increase in effective viscosity $\mu/\muf$ as a function of the volume fraction $\Phi$ for several different values of the Capillary numbers $\Ca=0.02(+)$, $0.1(\times)$, $0.2(\ast)$, $0.4(\boxdot)$, and $2(\bigtriangleup)$.  All the cases have $K=1$. For comparison we also plot  the same data for rigid particles~\cite{picano_breugem_mitra_brandt_2013a}, ($\Ca=0(\bullet)$). The inset shows the same data re-plotted as a function of $\Ca$ for different $\Phi \approx 0.0016$(green), $0.11$(blue), $0.22$(magenta), and $0.33$(red). (bottom) The same data re-plotted as a function of effective volume fraction $\phie$ collapses to an universal function given by the Eiler fit, \Eq{eq:Eiler} with $\phim = 0.6$ and $B = 1.7\/$. The horizontal error-bars show the standard deviation of the effective volume fraction. In the figure we also show the fitted data for three more cases at $\Phi=0.11$, $\Ca=0.2$, and viscosity ratio $K=0.01$(black), $0.1$(brown), and $10$(orange). The interested reader is referred to the Supplemental material for the discussion of the effect of $K$ on the effective viscosity.}
  \label{fig:viscFG}
\end{figure}

We use direct numerical simulations (DNS) of deformable spheres in plane Couette flow to calculate $\mu(\Phi,\Ca)$, for a wide range of  $\Phi$ (up to $\approx 33\%$) and $\Ca$ ($0.02-2$). We find that $\mu$ increases as $\Phi$ increases and decreases as $\Ca$ increases, \ie we find shear-thinning due to deformability. More importantly,  the function $\mu(\Phi,\Ca)$ collapses to an universal function, $\mu(\phie)$, see \Fig{fig:viscFG}, with a {\it reduced effective volume fraction} $\phie(\Phi,\Ca)$. Here $\phie$ is not a fit-parameter, but found independently from the shape of the deformed particles in the suspensions. Remarkably, the function $\mu(\phie)$ is well described by the Eilers fit, \Eq{eq:Eiler}. This demonstrates a striking universality of complex fluids: the Eilers fit  works for non-Brownian inertialess suspensions of rigid objects, suspensions at moderate $\Rey$  and also for non-Brownian suspensions of deformable objects, provided one uses $\phie$ instead of $\Phi$.
\begin{figure}
  \centering
  \includegraphics[width=0.45\textwidth]{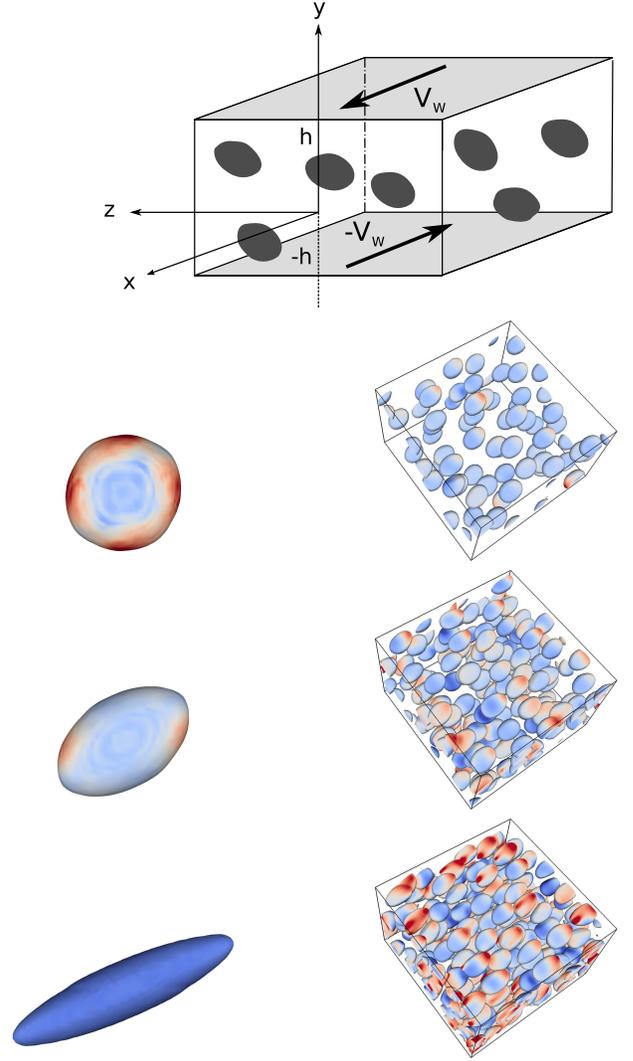}
  \caption{(top) Sketch of the channel geometry. The top and bottom walls, located at $y=\pm h$, move with opposite velocities $\pm V_{\rm w}$ in the stream-wise $x$ direction. (left) Shape of the deformed particle for lowest $\Phi$ (just single object in the computational box)  for three different Capillary numbers: $\Ca=0.02$, $0.2$, and $2$. (right) Shape of the deformed particles at $\Ca=0.2$ for three different volume fractions: $\Phi=0.11$, $0.22$, and $0.33$. The intensity of color shows $B^{12}$. In all our simulations we use $\Rey = 0.1$ with several different values of $\Phi \approx 0.0016$, $0.11$, $0.22$, and $0.33$, and $\Ca = 0.02$, $0.1$, $0.2$, $0.4$, and $2$. We use the viscosity ratio $\mus/\muf=1$ for all our simulations except for three more cases with $\phi=0.11$ and $\Ca=0.2$ where  $\mus/\muf=0.01$, $0.1$, and $10$.}
  \label{fig:sketch}
\end{figure}

We perform DNS in the plane Couette geometry -- see \Fig{fig:sketch} for a sketch of our computational box. The  {\it deformable} spheres suspended in the Newtonian fluid are modeled with a two-phase approach: the local volume fraction is denoted by $\phi$ \ie $\phi=1$ inside the viscoelastic solid phase and $\phi=0$ in the fluid phase, with a sharp boundary in between; hence $\Phi = \bra{\phi}$ where $\bra{\cdot}$ denotes volume average~\footnote{A similar approach gives rise to the Navier--Stokes--Cahn--Hilliard equations in binary fluids.}. The incompressible Navier--Stokes equations are solved everywhere for a monolithic velocity field~\cite{tryggvason_sussman_hussaini_2007a, takeuchi_yuki_ueyama_kajishima_2010a, quintard_whitaker_1994b}, $\uu$, and a stress tensor, $\sigma^{ij}$, given by 
\begin{subequations}
\label{eq:phi-stress}
\begin{align}
\sigma^{ij} &= \phi\sigmas^{ij} + \left(1-\phi\right)\sigmaf^{ij} \/, \\
\sigmas^{ij} &= -p \delta^{ij} +  2\mus D^{ij} + G B^{ij}\/, \\
\sigmaf^{ij} &= -p \delta^{ij} +  2\muf D^{ij}  \/.
\end{align}
\end{subequations}
Here the suffixes ${\rm f}$ and ${\rm s}$ indicate the fluid and solid phase, $D^{ij} \equiv (1/2)(\partial^iu^j + \partial^ju^i)$ the rate-of-strain tensor, $p$ the pressure and $\delta^{ij}$  the Kronecker delta. Clearly the fluid phase is a Newtonian one with dynamic viscosity $\muf$ and the solid phase is both viscous ($\mus$) and hyper-elastic with left Cauchy-Green tensor $B^{ij}$. Both $\phi$ and $B^{ij}$ are conserved quantities advected by the local velocity $\uu$.     

The dynamical equations are solved using a second order finite-difference scheme in space and third order Runge-Kutta scheme in time. The pressure is obtained by solving the Poisson equation using Fourier transforms. We use a Cartesian uniform mesh in a rectangular box of size $16R \times 10R \times 16R$, with 16 grid points per particle radius $R$. Periodic boundary conditions are imposed in the stream-wise $x$ and span-wise $z$ directions and no-slip  conditions  at the walls located at $y=-h$ and $y=h$, with $y$ the wall-normal direction, which move in opposite direction with constant stream-wise velocity $\pm V_{\rm w} = h \gdot$. We have validated our code by reproducing the results of Ref.~\cite{sugiyama_ii_takeuchi_takagi_matsumoto_2011a}, and details of our implementation can be found in Ref.~\cite{rosti_brandt_2017a}. We have checked that doubling the resolution in all directions results in an insignificant (less than $0.5\%$) change in the results. Also, the size of the domain has been chosen sufficiently large to avoid confinement effects \cite{picano_breugem_mitra_brandt_2013a, fornari_brandt_chaudhuri_lopez_mitra_picano_2016a}. The list of parameters investigated are given in the caption of \Fig{fig:sketch}. 

\begin{figure}[]
  \centering
  \includegraphics[width=0.45\textwidth]{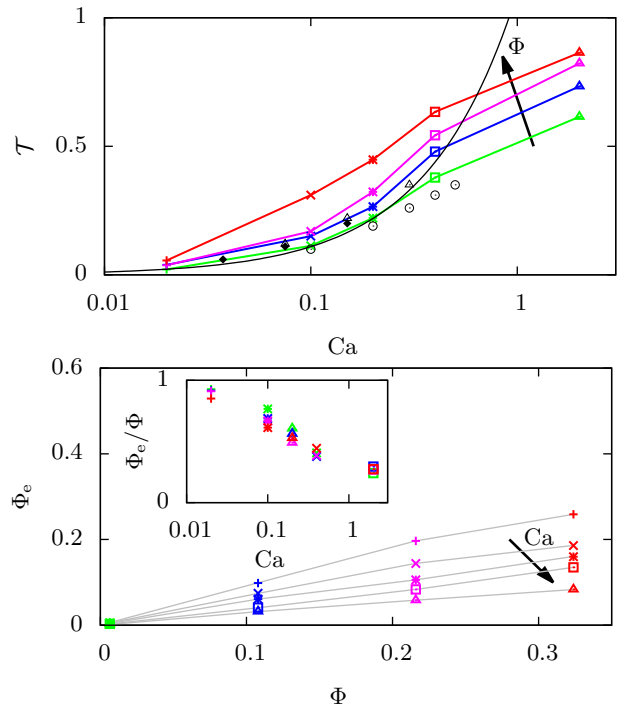}
  \caption{(top) The Taylor deformation parameter $\Tay$~\Eq{eq:Taylor} as a function of Capillary number $\Ca=0.02(+)$, $0.1(\times)$, $0.2(\ast)$, $0.4(\boxdot)$, and $2(\bigtriangleup)$ for $\Phi \approx 0.0016$(green), $0.11$(blue), $0.22$(magenta), and $0.33$(red). All the cases have $K=1$. The black solid line shows the result of the perturbative calculation of  Ref~\cite{choi_schowalter_1975a} expected to hold for small $\Ca$ and $\Phi$. The black symbols are numerical results from the literature for a single particle in a box. In particular, the triangle and rhombus are the results from Refs.~\cite{pozrikidis_1995a,eggleton_popel_1998a} which were calculated for $\Phi\approx0.06$, while the circle the $2$D simulation from Ref.~\cite{ii_sugiyama_takeuchi_takagi_matsumoto_2011a} with $\Phi\approx0.05$. (bottom) The effective volume fraction $\phie$ as a function of the volume fraction $\Phi$. The inset shows $\phie/\Phi$ as a function of the Capillary number $\Ca$, with logarithmic scale for the $x$-axis.}
  \label{fig:defG}
\end{figure}

We first run a set of simulations with the smallest $\Phi\approx0.0016$ which corresponds to one sphere in the computational volume. After the transients die out, the sphere deforms to approximately an ellipsoid. Examples are shown in \Fig{fig:sketch}, left column, for three different values of $\Ca$. We characterize these shapes by the Taylor parameter~\cite{taylor_1932a}
\begin{equation}
\label{eq:Taylor}
\Tay=\dfrac{b-a}{b+a}\/,
\end{equation}
where $b$ and $a$ are the lengths of the semi-major and semi-minor axis in the shear $xy$ plane. For higher values of $\Phi$, we start our DNS with the spheres randomly distributed in the computational domain and then wait until $T_{\rm tr} = 20/\gdot$ to reach statistical stationary state \cite{srivastava_malipeddi_sarkar_2016a}. Typical snapshots of the suspensions are shown in \Fig{fig:sketch}, right column, for three different values of $\Phi$. We calculate $\Tay$ by averaging over all the ellipsoids and plot $\Tay(\Phi,\Ca)$ in \Fig{fig:defG}. We also show the results of the perturbative analysis of Ref.~\cite{choi_schowalter_1975a}, which, as expected, agrees with our results at small $\Ca$ and small $\Phi$, and results from the numerical simulations of single particles in a box in Refs.~\cite{pozrikidis_1995a, eggleton_popel_1998a, ii_sugiyama_takeuchi_takagi_matsumoto_2011a}. 

We calculate the effective viscosity,  $\mu(\Phi,\Ca)$, as the ratio between the shear stress at the walls and $\gdot$. The effective viscosity $\mu$, normalized by $\muf$, as a function of $\Phi$ for several different values of $\Ca$ and as a function of $\Ca$ for several different values of $\Phi$ is shown in \Fig{fig:viscFG}. Clearly, for a fixed $\Ca$, the effective viscosity increases with $\Phi$, whereas for a fixed $\Phi$, the effective viscosity decreases as the Capillary number increases. The increase of $\Ca$ can, on one hand, be interpreted as a decrease in $G$ (with $\gdot$ and $\muf$ held constant) \ie $\mu$ decreases as the spheres become more deformable. On the other hand, the increase of $\Ca$ can be interpreted as an increase of $\gdot$ (with $G$ and $\muf$ held constant) consequently  $\mu$ decreases as $\gdot$ increases, \ie we observe shear-thinning. This latter interpretation is valid only when the inertial effects remain vanishingly small. This is consistent with earlier studies~\cite{pal_2003a,srivastava_malipeddi_sarkar_2016a} for small $\Ca$ and $\Phi$ (for demonstration see the supplemental material).

At constant $\Phi$, as $\Ca$ increases, $\Tay$ increases and the spheres become approximately prolate spheroids aligned with the shear directions \cite{matsunaga_imai_yamaguchi_ishikawa_2016a}. This suggests that the shear-thinning (decrease in $\mu$) with increasing $\Ca$ can be interpreted in terms of a decrease in the {\it effective} volume fraction $\phie$, a concept successfully used in the past for suspensions with different properties, such as charged colloidal particles, fiber and platelets suspensions, polyelectrolyte solutions \cite{mewis_frith_strivens_russel_1989a, frith_dhaene_buscall_mewis_1996a, quemada_1998a, mewis_wagner_2012a}. Here, we define it by $\phie = (4\pi/3)\bra{a}^3/\V$ where $\bra{a}$ is the mean semi-minor axis of all the particles calculated from the DNS and $\V$ the total volume of the computational box. We use the variance of $a$ to estimate of the error in $\phie$. The choice of using the minor axis is different from what done in previous works for fiber suspensions \cite{batchelor_1971a, kerekes_2006a, lundell_soderberg_alfredsson_2011a}, where the major axis is usually considered. This is motivated by the fact that in our case the particles are not tumbling and are approximately aligned with the mean shear direction, thus, what matters is the dimension in the direction normal to the mean shear, \ie the minor axis. The reduced volume fraction $\phie$ increases with $\Phi$ and decreases with $\Ca$, see \subfig{fig:defG}{b}. Furthermore, we find that $\phie/\Phi$ is a function of $\Ca$ alone, see inset of \subfig{fig:defG}{b}, a finding useful for future modeling. This brings us to the central result of this letter in \subfig{fig:viscFG}{b}: the effective viscosity $\mu/\muf$ plotted as a function of $\phie$ for all the different cases collapses to a universal function, \ie we have shown that the effect of the deformability of the particles can be included into the effective viscosity of the suspension as follow
\begin{equation}
\mu/\muf=\mathcal{F} \left[ \phie \left( \Phi, \Ca \right) \right],
\end{equation}
where $\phie$ is the effective volume fraction encoding the deformation, and $\mathcal{F}$ an universal function. Remarkably, we find that the Eilers fit, \Eq{eq:Eiler}, with $\Phi$ replaced by $\phie$  provides a good description of this universal function. Data for four different values of the viscosity ratio $K$ are included in \subfig{fig:viscFG}{b}, which also collapse to the universal Eilers fit. As shown in the Supplemental material, we find that $\mu/\muf$ depends weakly on the viscosity ratio, $K$. The data from another recent DNS~\cite{matsunaga_imai_yamaguchi_ishikawa_2016a} of fluid-filled deformable capsules can also be collapsed to the universal Eilers fit; see Supplemental material.
\begin{figure}[]
  \centering
  \includegraphics[width=0.45\textwidth]{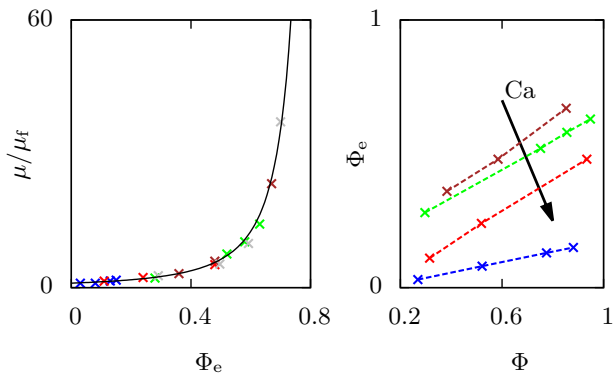}
  \caption{(left) The effective viscosity of suspensions of RBCs for different deformabilities and viscosity ratio plotted as a function of the effective volume fraction collapses to the Eilers fit. The data are obtained from Fig. 5 of Ref.~\cite{dintenfass_1968a}. The brown points correspond to normal RBCs in saline, while the blue, red and green ones to RBCs in dextran solutions of viscosities $3.2$, $11$ and $67$ centipoises, respectively. The grey symbols are the rigid RBCs treated with acetylaldehyde. (right) The effective volume fraction $\phie$ as a function of $\Phi$ necessary to obtain the collapse in the left panel.}
  \label{fig:RBC}
\end{figure}

Finally, we demonstrate how robust our results are by applying them to {\it experimental} data~\cite{dintenfass_1968a}  on viscosity of suspensions of Red Blood Cells (RBCs) -- to distinguish such suspension from blood, which is a more complex system, we call~\cite{fedosov_pan_caswell_gompper_karniadakis_2011a} them erythrocite suspensions~(ES). Although several experiments have measured the effective viscosity of erythrocite--suspensions under a range of volume fractions and shear-rates~\cite[see, e.g.,][for recent review of numerical and experimental results.]{winkler_fedosov_gompper_2014a, yazdani_li_karniadakis_2016a} only Ref.~\cite{dintenfass_1968a} measured effective viscosity at four different capillary numbers too by changing the viscosity, $\muf$, of the solvent~\footnote{The viscosity ratio, $K$, changes too but our results show that the $\mu$ does not depend strongly on $K$.} and compared it against one rigid case obtained by treating the RBCs with acetaldehyde. To apply our result to these data we first fit the Eilers formula to the case of the hard RBCs, obtaining $B=1.25$ and a maximum packing fraction $\phim = 0.88$, as the undeformed shape of the RBCs is not spherical but disk-like~\cite{mueller_llewellin_mader_2009a}. With these changes we find that the viscosity of RBCs can be collapsed to the Eilers fit as shown in \subfig{fig:RBC}{a}. As necessary condition for this collapse we obtain the dependence of the effective volume fraction $\phie$ with $\Phi$ shown in \subfig{fig:RBC}{b}. The curves are approximately linear, and  decrease with $\Ca$ for a fixed $\Phi$, which is similar to what we have obtained from our simulations, \subfig{fig:defG}{b}. As various diseases, including malaria and sickle cell anemia, increase the deformability of RBCs, our results suggests the intriguing possibility that it may also be possible to use our method to model the change in effective viscosity of blood in such cases~\footnote{This applies to macro-rheology of blood not micro-rheology which deals with blood flows in capillaries of sizes close the size of the RBCs themselves.}. 

To conclude, our simulations show that a suspension of deformable incompressible spheres in a Newtonian fluid displays shear-thinning and that this can be understood in terms of a reduction of the effective volume fraction occupied by the suspended spheres due to their deformation. Considered in conjunction with earlier results~\cite{picano_breugem_mitra_brandt_2013a} we find that the Eilers fit used with the concept of effective volume fraction is a surprising powerful too to interpret rheological data. In other words, the suspension dynamics is mainly determined by excluded volume effects for non-Brownian suspensions of rigid and deformable particles, the former also in the weakly inertial regime. A word of caution though, not all aspects of non-Brownian suspensions can be described by an effective viscosity, e.g., laminar to turbulent transition in a suspension is qualitatively different from that of a Newtonian fluid~\cite{lashgari_picano_breugem_brandt_2014a}. In view of our initial success in interpreting existing rheological measurement of suspension of RBCs we suggest a systematic experimental investigation of suspensions of cells and capsules with different deformability.  

\section*{Acknowledgment}
The work of MER and LB was supported by the European Research Council grant no. ERC-2013-CoG-616186, TRITOS and by the Swedish Research Council (grant no. VR 2014-5001). DM is supported by grants from the Swedish Research Council (grant no. 638-2013-9243 and 2016-05225). The authors acknowledge computer time provided by SNIC (Swedish National Infrastructure for Computing).

\bibliography{../../../../Articles/bibliography.bib}

\newpage
~
\newpage
\beginsupplement
\section*{Supplemental material}
\begin{figure}[h]
 \centering
  \includegraphics[width=0.46\textwidth]{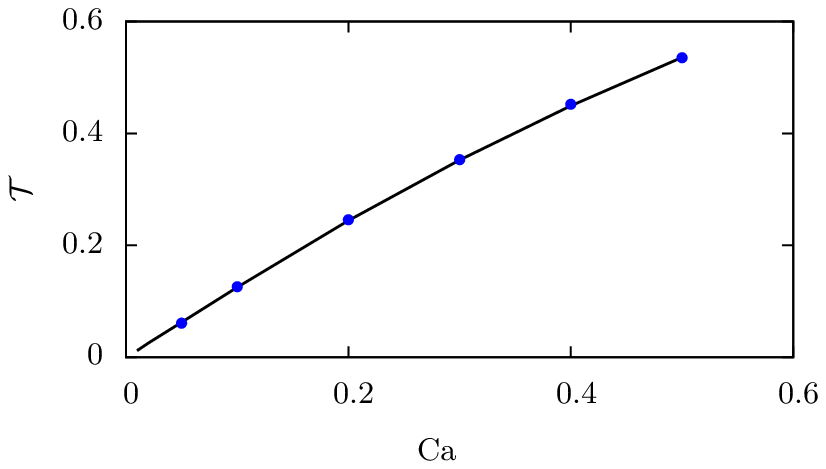}
\caption{The Taylor deformation parameter $\Tay$~\Eq{eq:Taylor}
as a function of Capillary number $\Ca$ for dilute suspensions with $K=0$. The black solid line shows the numerical result by Ref~\cite{villone_hulsen_anderson_maffettone_2014a}, while the blue symbols are the results of our simulations.}
\label{fig:valM}
\end{figure}
\Fig{fig:valM} shows the validation of our numerical method with recent results from the literature \cite{villone_hulsen_anderson_maffettone_2014a}; in particular, we show the Taylor deformation parameter $\Tay$~\Eq{eq:Taylor} as a function of Capillary number $\Ca$ for dilute suspensions with $K=0$, and we find a very good agreement.

\begin{figure}[h]
 \centering
  \includegraphics[width=0.45\textwidth]{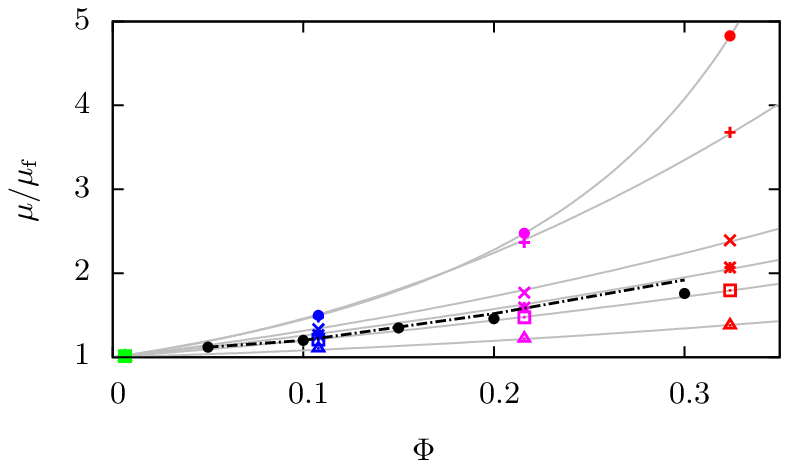}
\caption{ The fractional increase in effective viscosity $\mu/\muf$ as a function of the volume fraction 
$\Phi$ for several different values of the Capillary numbers $\Ca=0.02(+)$, $0.1(\times)$, $0.2(\ast)$, $0.4(\boxdot)$, and $2(\bigtriangleup)$, same as Fig. (1). For comparison, we report also Pal's empirical relation \cite{pal_2003a} with the dash-dotted black line and Srivastava \etal results \cite{srivastava_malipeddi_sarkar_2016a} with the black circles for $Ca=0.15$.}
\label{fig:viscF}
\end{figure}
The effective viscosity  $\mu(\Phi,\Ca)$ as a function of $\Phi$ for several 
different values of $\Ca$ is shown in \Fig{fig:viscF}. The figure is the same as \Fig{fig:viscFG}, with the addition of reference data from Pal's empirical relation \cite{pal_2003a}, plotted with the dash-dotted line, and the numerical results by Srivastava \etal \cite{srivastava_malipeddi_sarkar_2016a}, plotted with the black circles.

\begin{figure}
  \centering
  \vspace{0.5cm}
  \includegraphics[width=0.445\textwidth]{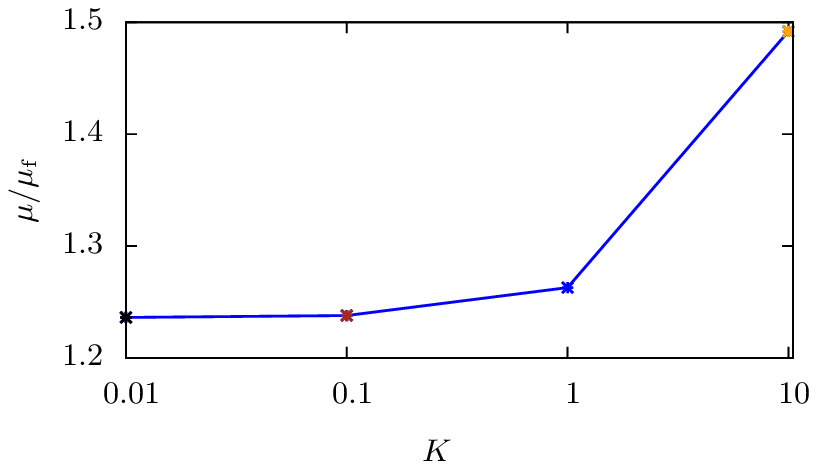}
  \caption{Effective viscosity $\mu/\muf$ as a function of the solid to fluid viscosity ratio, $K$, in a logarithmic scale. The Capillary number is fixed to $\Ca=0.2$ and the volume fraction to $\Phi=0.1$.}
  \label{fig:viscV}
\end{figure}
In \Fig{fig:viscV}, we display $\mu/\muf$ as a function of the ratio of the two viscosities $K \equiv \mus/\muf$ for fixed $\Ca=0.2$ and $\Phi=0.1$. By changing  $K$ by a factor of $1000$, $\mu/\muf$ varies only by a factor of $1.2$, \ie the effective viscosity has a weak dependence on $K$, in agreement with the results in~\cite{pal_2003a}. Our results suggest that the effective viscosity is mainly determined by $\Ca$ (\Fig{fig:viscF}) and only weakly by the viscosity ratio $K$ (\Fig{fig:viscV}).

\begin{figure}
  \centering
  \includegraphics[width=0.45\textwidth]{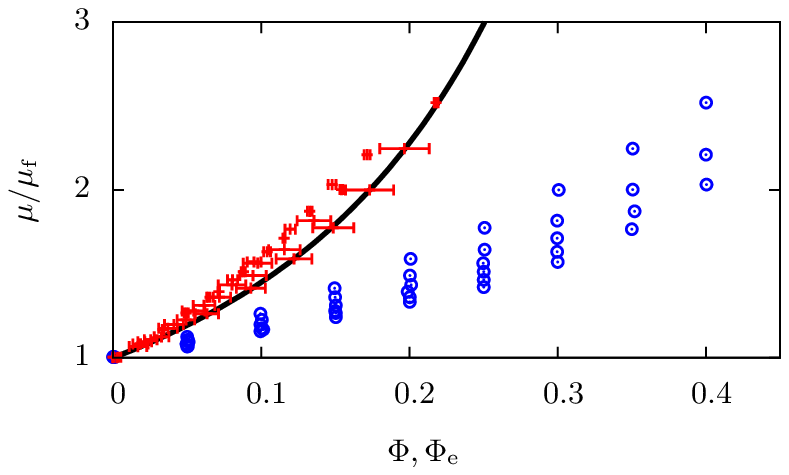}
  \caption{Effective viscosity $\mu/\muf$ as a function of the volume fraction $\Phi$ (blue symbols) and as a function of the effective volume fraction $\phie$ (red symbols) for the results by Ref.~\cite{matsunaga_imai_yamaguchi_ishikawa_2016a} of a suspension of spherical capsules. The error-bar is computed based on the standard deviation of the Taylor parameter $\Tay$.}
  \label{fig:viscVAL}
\end{figure}
In \Fig{fig:viscVAL}, we display the validity of our fit based on the results in Ref.~\cite{matsunaga_imai_yamaguchi_ishikawa_2016a} for a suspension of spherical capsules. The blue symbols are the original data, \ie $\mu/\muf$ as a function of the volume fraction $\Phi$, while the red ones are the fitted ones, \ie $\mu/\muf$ as a function of the effective volume fraction $\phie$ computed from the Taylor parameters reported in the manuscript.

\end{document}